\documentclass[prb, twocolumn, preprintnumbers, amsmath, amssymb,
  nofootinbib, floatfix, superscriptaddress]{revtex4}

\usepackage{graphicx,bm,xcolor, bbold}
\usepackage{enumerate}
\usepackage{graphicx,bm}
\usepackage[normalem]{ulem} 
\usepackage[breaklinks]{hyperref}

\makeatletter
\def\graphicscale{\twocolumn@sw{0.33}{0.4}}
\def\graphicthreescale{\twocolumn@sw{0.3}{0.4}}

\begin{document}

\title{Interplay between short-range and critical long-range fluctuations
  in \\ the out-of-equilibrium behavior of the
  particle density at quantum transitions}

\author{Davide Rossini}
\affiliation{Dipartimento di Fisica dell'Universit\`a di Pisa and INFN,
  Largo Pontecorvo 3, I-56127 Pisa, Italy}

\author{Ettore Vicari}
\affiliation{Dipartimento di Fisica dell'Universit\`a di Pisa,
  Largo Pontecorvo 3, I-56127 Pisa, Italy}

\date{\today}

\begin{abstract}
  We address the equilibrium and out-of-equilibrium behavior of the
  particle density in many-body systems undergoing quantum transitions
  driven by the chemical potential $\mu$.  They originate from a
  nontrivial interplay between noncritical short-range and critical
  long-range quantum fluctuations.  As a paradigmatic model we
  consider the one-dimensional fermionic Kitaev chain, for which very
  accurate numerical studies can be performed, up to $O(10^4)$ 
  sites.  The search for dynamic scaling behaviors of the particle
  density is complicated by the fact that its equilibrium (ground
  state) behavior is dominated by short-range fluctuations, giving
  rise to regular background terms and peculiar logarithmic terms from
  resonances between renormalization-group perturbations associated
  with the {\em energy} and {\em identity} operator families within
  the conformal field theory. To study these issues, we focus on two
  dynamic protocols, either instantaneous quenches or quasi-adiabatic
  changes of $\mu$ to the critical value $\mu_c$, unveiling
  out-of-equilibrium scaling behaviors of the particle density, which
  arise from the critical modes, within a dynamic finite-size scaling
  framework.
\end{abstract}

\maketitle

% ========================= BODY =========================

\section{Introduction}
\label{intro}

At continuous quantum transitions in the zero-temperature
limit~\cite{Sachdev-book, RV-21}, the behavior of some interesting
observables displays a nontrivial interplay of contributions from
(noncritical) short-range and (critical) long-range quantum
fluctuations, which can be hardly disentangled.  This is the case of
the particle density at quantum transitions driven by the chemical
potential, whose critical scaling behavior gets hidden by
contributions coming from the regular (short-range) term of the
free-energy density~\cite{CPV-14}.  This phenomenon may be
considered as the quantum counterpart of the analogous interplay
between short-range and critical fluctuations at thermal classical
transitions, where the energy density at the transition point is
dominated by a regular term arising from short-range fluctuations (or
mixing with the identity operator), while the critical scaling terms
are only subleading~\cite{WK-74, Fisher-74, Wegner-76, Privman-90,
  SS-00, PV-02}.  Therefore, in equilibrium conditions, the energy
density in classical transitions and the particle density at quantum
transitions (driven by the chemical potential) generally display
nonuniversal leading behaviors at the transition point, while critical
scaling behaviors are relatively suppressed. For this reason, they
are not considered as optimal observables to probe the universal
features at criticality.

However, despite this equilibrium behavior, some numerical analyses of
out-of-equilibrium behaviors of three-dimensional $N$-vector models at
classical (thermal) transitions have provided evidence of a peculiar
out-of-equilibrium scaling of the energy density~\cite{PV-24}.  This
has been observed along the critical relaxational flow arising from
instantaneous quenches of the temperature to the critical point, under
a purely relaxational dynamics that leads to the asymptotic large-time
thermalization~\cite{HH-77, Ma-book}, starting from equilibrium
conditions.  The out-of-equilibrium finite-size scaling (FSS) behavior
of the energy density (after subtracting its asymptotic value at the
critical point) can be expressed in terms of a rescaled time variable
$\Theta=t/L^z$ (where $t$ is the time from the quench, $z$ is the
dynamic exponent associated with the relaxational dynamics, and $L$ is
the system size), analogously to other observables that exhibit
scaling at equilibrium. This may be related to the fact that the
dynamics of short-range fluctuations is characterized by a
significantly shorter time scale with respect to the diverging time
scale of critical modes.  However, the absence of a scaling behavior
at equilibrium, thus at the starting point of the quenching protocol,
leaves a noticeable imprinting in the dynamic scaling behavior as a
function of $\Theta$, since scaling functions show a peculiar
power-law singularity in the $\Theta\to 0$ limit, and in particular a
power-law divergence when the specific-heat exponent is
negative~\cite{PV-24}.

In this paper, we investigate whether analogous phenomena emerge in
many-body systems at quantum transitions driven by the chemical
potential, focusing on the out-of-equilibrium behavior of the particle
density arising from instantaneous and slow quenches of the
chemical-potential parameter $\mu$ to the critical point.
Although the classical-to-quantum mapping does not apply, since its
applicability is restricted to equilibrium conditions,
out-of-equilibrium scaling behaviors similar to those at thermal
transitions have been also observed at quantum transitions for
standard observables showing asymptotic equilibrium scaling behaviors,
such as the correlation function of the order-parameter operator (see,
e.g., Refs.~\onlinecite{RV-21, ZDZ-05, Dziarmaga-05, DOV-09, GGP-10,
  GZHF-10, Dziarmaga-10, PSSV-11, CEGS-12, KCH-12, FDGZ-16, CC-16,
  Biroli-16, V-18, PRV-18, NRV-19-wo, NRV-19, RV-19, RDZ-19, PRV-20,
  RV-20-kz, TV-22, DV-23, TV-23}).  Again, like classical transitions,
the extension to observables dominated by regular terms (mixing with
the identity operator) at equilibrium is not straightforward, because
of the nontrivial competition between contributions arising from
noncritical short-range and critical long-range fluctuations.
Moreover, quantum scenarios may substantially differ from those
observed at classical transitions under relaxational dynamics,
essentially because the quantum dynamics of isolated systems is
qualitatively different, being unitary and energy conserving. Thus,
further interesting features may emerge, with respect to those
observed along critical relaxational flows at classical transitions,
always characterized by an asymptotic large-time thermalization.

To study this issue, we focus on the one-dimensional Kitaev
chain~\cite{Kitaev-01}, as a paradigmatic model for quantum
transitions driven by the chemical potential.  Its equilibrium
particle density at the quantum critical point is dominated by a
standard regular term~\cite{CPV-14, RV-21} and also logarithmic terms
arising from peculiar resonances of the renormalization-group (RG)
weights of RG operators~\cite{Wegner-76, CPV-14}, belonging to the
{\em energy} and {\em identity} conformal families within a
conformal-field-theory (CFT) framework (see, e.g.,
Ref.~\onlinecite{CHPV-02}).  We recall that this resonance phenomenon
is responsible for the logarithmic divergence of the specific heat at
the classical thermal transitions belonging to the two-dimensional
(2D) Ising universality class~\cite{Wegner-76}.
We consider two relatively simple dynamic protocols: (i) quantum
quenches, where $\mu$ is instantaneously changed to the critical-point
value $\mu_c$, and (ii) quasi-adiabatic variations of the
time-dependent parameter $\mu(t)$ from $\mu\neq\mu_c$ to $\mu_c$,
starting in both cases from equilibrium (ground-state) conditions.  We
are able to uncover, and characterize, the emergence of
out-of-equilibrium scaling behaviors of the particle density arising
from the critical modes, within a dynamic FSS framework.

Our numerical study shows that the particle density (after subtracting
its large-volume critical-point value) develops a dynamic scaling
behavior along the post-quench quantum evolution, similar to that
observed along critical relaxational flows at classical transitions.
The corresponding scaling functions are characterized by a logarithmic
divergence in the $\Theta\to 0$ limit, where $\Theta\sim t/\tau$ is
the rescaled time with respect to the time scale of the critical modes
[$\tau\sim \Delta_c(L)^{-1}\sim L^z$, where $\Delta_c(L)$ is the gap
  at the critical point and $z=1$ is the dynamic exponent], somehow
reflecting the fact that the subtracted particle density does not have
an asymptotic equilibrium scaling behavior.  Moreover, the scaling
functions show some further logarithmic divergence at finite $\Theta$,
related to revival quantum phenomena. We also show that an
out-of-equilibrium FSS behavior emerges along protocols entailing
slow quasi-adiabatic changes of the Hamiltonian parameters, once 
subtracting the corresponding equilibrium particle density at the
instantaneous value of $\mu(t)$.

The paper is organized as follows. In Sec.~\ref{kitaevmod} we present
the fermionic Kitaev wire, providing a paradigmatic model which
undergoes a quantum transition driven by the chemical potential.  In
Sec.~\ref{sec:scaling} we discuss the equilibrium behavior of the
particle density at the quantum transition, which does not show a
universal asymptotic scaling. In Sec.~\ref{outeq} we outline the
dynamic quench protocol and discuss the out-of-equilibrium behavior
within a dynamic FSS framework; moreover, we present numerical results
showing that the particle density develop an out-of-equilibrium FSS
along the post-quench quantum evolution, with peculiar singularities
of its FSS functions. We also derive the corresponding
out-of-equilibrium scaling behavior in the thermodynamic limit.
In Sec.~\ref{KZprot} we extend our analysis to
quasi-adiabatic dynamic protocols entailing slow changes of the
chemical potential. Finally, in Sec.~\ref{conclu} we summarize and
draw our conclusions.

\section{The fermionic Kitaev chain}
\label{kitaevmod}

As a paradigmatic many-body system undergoing a continuous quantum
transition driven by the chemical potential, we consider a fermionic
Kitaev wire of size $L$ (number of sites), whose quantum unitary
dynamics is driven by the Hamiltonian~\cite{Kitaev-01}
\begin{equation}
  \hat H = - J \sum_{x} \big( \hat c_x^\dagger \hat
  c_{x+1}^{\phantom\dagger} + \gamma \hat c_x^\dagger \hat
  c_{x+1}^\dagger+{\rm h.c.}  \big) - \mu \hat{N},
  \label{kitaev}
\end{equation}
where $\hat c_x$ is the fermionic annihilation operator associated
with the $x$th site of the chain ($x=1, \ldots, L$), while $\hat{N}$
is the particle-number operator:
\begin{equation}
  \hat{N} = \sum_x \hat{n}_x, \qquad \hat n_x=\hat c_x^\dagger \hat
  c_x^{\phantom\dagger}.
  \label{partnumb}
\end{equation}
The Hamiltonian parameter $\mu$ denotes the chemical potential,
while $\gamma>0$ controls the relative strength of the terms which
do not conserve the fermionic number.  In the following, we fix the
energy scale by assuming $J=1$ and also set the Planck
and Boltzmann constants $\hslash = k_B = 1$.

The Hamiltonian~\eqref{kitaev} can be straightforwardly diagonalized
into~\cite{LSM-61, Katsura-62, Pfeuty-70, BG-85}
\begin{equation}
  \hat H = \sum_k E(k) \left( \hat a^\dagger_k \hat a_k -
  \tfrac12 \right),
  \label{H-harmonic}
\end{equation}
where $\hat a_k$ are new fermionic annihilation operators, which are
obtained through a suitable linear transformation of the original
$\hat c_x$ operators, and
\begin{equation}
  E(k) = 2 \sqrt{
    (\mu/2)^2 + \gamma^2 + \mu \cos k + (1 - \gamma^2) \cos^2 k}
  \label{Ek-XY}
\end{equation}
is their dispersion relation.  In a finite system, the set of $k$
values to summed over, as well as the allowed states, depend on the
boundary conditions.

By means of the Jordan-Wigner transformation (see, e.g.,
Ref.~\onlinecite{Sachdev-book}), the Hamiltonian~\eqref{kitaev}
can be also mapped into a so-called quantum $XY$ chain:
\begin{equation}
  \hat H_{XY} \!= \!- \! \sum_{x} \! \left[ \frac{1 + \gamma}{2} \hat
    \sigma^{(1)}_x \hat \sigma^{(1)}_{x+1} + \frac{1 - \gamma}{2}
    \hat \sigma^{(2)}_x \hat \sigma^{(2)}_{x+1} + g \hat
  \sigma^{(3)}_x\right]\! ,
  \label{XYchain}
\end{equation}
where $\hat \sigma^{(k)}_x$ are the spin-$1/2$ Pauli matrices
($k=1,2,3$). In particular, $\hat\sigma^{(3)}_x = 1 - 2 \hat{n}_x$,
thus $g=-\mu/2$.  However, although the bulk behaviors of the two
models in the infinite-volume limit (and thus their phase diagram) are
analogous, some finite-size features may differ significantly.  For
example, the nonlocal Jordan-Wigner transformation of the $XY$
chain~\eqref{XYchain} with periodic and antiperiodic boundary
conditions does not map into the fermionic model~\eqref{kitaev} with
analogous boundary conditions. Indeed further considerations apply,
leading to a less straightforward correspondence, which also depends
on the parity of the particle-number eigenvalue (see, e.g.,
Refs.~\onlinecite{Katsura-62, Pfeuty-70, CPV-14}).  Therefore, the Kitaev
quantum wire and the quantum $XY$ chain cannot be considered as
completely equivalent.  However, they both undergo a continuous
quantum transition, respectively at $\mu = \mu_c = -2$ or at $g = g_c
= 1$, independently of the parameter $\gamma$.  To simplify the
notation, we define the deviation of the relevant parameter $\mu$ from
its critical value as
\begin{equation}
  w \equiv {\mu_c - \mu \over 2} = g - g_c , \qquad (\mu_c=-2, \; g_c=1).
  \label{wdef}
\end{equation}

In the following we consider the fermionic Kitaev chains with
antiperiodic boundary conditions~\cite{NRV-19}, i.e. with $\hat{c}_x =
-\hat{c}_{L+x}$, which simplify computations of the equilibrium and
out-of-equilibrium FSS behaviors, restricting the momenta of the sum
in Eq.~\eqref{H-harmonic} to
\begin{equation}  
  k = \Big\{ \pm \frac{\pi}{L}(2n+1) \Big\} , \quad n=0,1,\ldots, L/2-1 .
  \label{kval}
\end{equation}
When considering antiperiodic boundary conditions, both phases
separated by the quantum transition at $\mu_c$ are gapped, i.e., the
degeneracy of the vacua for $\mu < \mu_c$ in the thermodynamic limit
(corresponding to the ordered phase of quantum $XY$ chains) is not
realized. The reason for such substantial difference resides in the
fact that the corresponding Hilbert space is restricted with respect
to that of the $XY$ chain, so that it is not possible to restore the
competition between the two vacua belonging to the
symmetric/antisymmetric sectors of the quantum $XY$
chain~\cite{Katsura-62, Kitaev-01, CPV-14, RV-21}.

\section{Equilibrium behavior of the particle density}
\label{sec:scaling}

\subsection{Quantum criticality}
\label{quacribeh}

The continuous transition at $\mu_c$ belongs to the 2D Ising
universality class~\cite{Sachdev-book, RV-21}, characterized by the
length-scale critical exponent $\nu=1$, related to the RG dimension
$y_w = 1/\nu=1$ of the Hamiltonian parameter $w$.  This implies that,
approaching the critical point $w\to 0$ at zero temperature, the
length scale $\xi$ of the critical quantum fluctuations diverges as
$\xi \sim |w|^{-\nu}$.  The temperature $T$ represents another
relevant RG perturbation at quantum transitions. At the critical point
$w=0$, the length scale increases as $\xi \sim T^{-1/z}$ with
decreasing $T$, where $z$ is dynamic exponent $z=1$ associated with
the unitary quantum dynamics within this universality class (it also
determines the power law $\Delta\sim\xi^{-z}$ of the vanishing gap
with increasing $\xi$).  Moreover, we mention that the RG dimension of
the fermionic operators $\hat c_x$ and $\hat c^\dagger_x$ at the
continuous quantum transition is $y_c = 1/2$, and that of the particle
density operator $\hat n_x$ is~\cite{Sachdev-book, RV-21}
\begin{equation}
  y_n = d+z-y_w = 2 - y_w = 1.
  \label{ynexp}
\end{equation}

The universal critical exponents enter the asymptotic power laws of
the quantum critical behavior as a function of the temperature $T$ and
the chemical potential $\mu$. However, the asymptotic critical
expansions associated with the 2D Ising universality class are also
characterized by the presence of logarithmic terms~\cite{Wegner-76,
  SS-00, Queiroz-00, Salas-01, ONGP-01, IH-02, CHPV-02, IH-09,
  CGNP-11, Izmailian-12, Izmailian-13}.  They arise from a resonance
between the {\em identity} operator of RG dimension 2 and the {\em
  energy} operator of RG dimension 1 within the corresponding 2D
conformal field theory (CFT) with central charge
$c=1/2$~\cite{CHPV-02}. In particular, such a resonance mechanism is
responsible for the leading logarithmic divergence of the specific
heat at the 2D Ising critical point.

Analogous logarithmic terms are found at the quantum critical point of
the quantum $XY$ chain, equivalent to the fermionic Kitaev wire in the
thermodynamic limit or for open boundary conditions. In particular,
they appear in the free-energy density
\begin{equation}
  F(w,T,\gamma) = - {T\over L}
  \ln \big[ {\rm Tr}\,e^{-\beta \hat H} \big],
  \qquad \beta = 1/T. \label{freeF}
\end{equation}
In the thermodynamic limit, i.e. when $L/\xi\to\infty$,
it can be written as~\cite{Katsura-62}
\begin{equation}
  F(w,T,\gamma) = - \int_{0}^{\pi} \, {dk\over 2\pi}
\, \left\{ E(k) + 2T \ln\Bigl[1 +
  e^{-\beta E(k)}\Bigr]\right\},
\label{fwtga}
\end{equation}
where $E(k)$ is reported in Eq.~\eqref{Ek-XY}.  At the leading order
in the critical limit, it behaves as~\cite{CPV-14} 
\begin{eqnarray}
  F(w,T,\gamma) &\approx& F_{\rm reg}(w,\gamma)
  + \label{FscalXY}\\ &+& {A \over 4\pi} u_w^2 \ln u_w^2
    - {2 A\over \pi}  u_t^2 \,f(u_w/u_t), \nonumber
\end{eqnarray}
where $F_{\rm reg}(w,\gamma)$ is a regular function at the critical
point, $u_w$ and $u_t$ are the scaling fields corresponding to the
relevant parameters $w$ and $T$, which are given by
\begin{equation}
u_w = {w\over A},\quad u_t = {T\over 2A},\quad A \equiv
  \sqrt{\gamma^2 + w},
  \label{umut}
\end{equation}
and $f(x)$ is a universal scaling function (apart from a trivial
factor and a normalization of the argument) given by
\begin{equation}
  f(x) = \int_0^\infty dz \, \ln \left(1 + e^{-\sqrt{x^2 + z^2}}\right).  
  \label{fxsca}
\end{equation}
Note that the first regular term of the expansion~\eqref{FscalXY} is
independent of $T$, as generally expected at quantum
transitions~\cite{CPV-14, RV-21}, and can be expanded in powers of $w$:
\begin{equation}  
  F_{\rm reg}(w,\gamma) = b_0(\gamma) + b_1(\gamma) \,w  + \ldots .
  \label{fregexp}
\end{equation}

\subsection{The particle density in the thermodynamic limit}
\label{partdensther}

The thermodynamic-limit behavior of the equilibrium particle density
$\varrho_e$ can be derived by differentiating the free-energy density
with respect to the chemical potential,
\begin{eqnarray}
  &&\varrho_e \equiv L^{-1} {\rm Tr} \,[\rho_G \, \hat{N}] 
  = -\partial_\mu F = \tfrac{1}{2} \partial_w F
  \label{rhobeh}\\ &&\;\;\approx \varrho_{\rm reg}(w,\gamma)+
        {B\over 2\pi} \left( u_w \ln u_w^2 + u_w\right) - {B \over \pi}
        u_t\, g(u_w/u_t), \nonumber
\end{eqnarray}
where $\rho_G$ is the density matrix associated with the Gibbs
distribution, 
\begin{equation}
  \rho_G = {e^{-\beta\hat{H}}\over {\rm Tr} [ e^{-\beta \hat H}]},
    \label{rhogibbs}
\end{equation}
and
\begin{equation}
  B=A \, \partial_w u_w = 1-w/(2A^2).
\end{equation}
In Eq.~\eqref{rhobeh} we
only kept the most relevant terms, and $g(x)$ is another scaling
function that can be easily derived from $f(x)$,
cf. Eq.~\eqref{fxsca}, i.e., $g(x)=\partial_x f(x)$. In the critical
limit $u_w, u_t\to 0$, keeping the ratio $u_w^{z\nu}/u_t=u_w/u_t$
fixed, the particle density is dominated by the contribution of the
regular term, which can be expanded as
\begin{equation}
  \varrho_{\rm reg}(w,\gamma) = a_0(\gamma) + a_1(\gamma)\, w + ...
  \label{rhoreg}
\end{equation}
where $a_i$ are nonuniversal constants depending on $\gamma$.  In
particular, $a_0=1/2 - 1/\pi$ for $\gamma=1$.

Actually, the regular background generally provides the leading
contribution to the free-energy density, and its derivative with
respect to the even relevant Hamiltonian parameter~\cite{RV-21},
analogous to $\mu$ in quantum transitions driven by the chemical
potential. In some cases one obtains an asymptotic scaling behavior in
the FSS limit by subtracting its the critical-point value, in
particular when the dynamic and length-scale critical exponents are
such that $d+z-2/\nu<0$ as it occurs in the 2D quantum Ising
model~\cite{RV-21}.  However, in the case of the fermionic Kitaev
wire, the particle-density deviation $D_e(w,T,\gamma)$ from its
critical-point value $\varrho_c(\gamma)$,
\begin{equation}
  D_e(w,T,\gamma) \equiv \varrho_e(w,T,\gamma) - \varrho_c(\gamma),
  \label{deltarhodef}
\end{equation}
where
\begin{equation}
  \varrho_c(\gamma) \equiv \varrho_e(0,0,\gamma)= a_0(\gamma),
  \label{rhoc}  
\end{equation}
tunrs out to be dominated by the logarithmic term arising from the
resonance between identity and energy operators, hiding the universal
scaling behavior~\cite{CPV-14, RV-21}
\begin{equation}
  \varrho_{\rm scal} \sim T^{y_n/z} g(u_w/u_t)
  \label{rhosca}
\end{equation}
[$y_n=1$ is the RG dimension of the particle-density operator
  $\hat{n}_x$, see Eq.~\eqref{ynexp}], which remains logarithmically
suppressed with respect to the leading term.

\subsection{Finite-size behavior of the particle density}
\label{fsspartbeh}

The above scaling behaviors can be straightforwardly extended to
finite-size systems, within a FSS framework (see, e.g.,
Refs.~\onlinecite{CPV-14, RV-21, FB-72, Barber-83, Privman-90,
  Cardy-editor, CHPV-02, PV-02}). Zero-temperature FSS ansatzs can be
obtained by introducing the lattice size $L$, rewriting
Eq.~\eqref{rhobeh} in terms of the scaling variable
\begin{equation}
  \Phi = u_w(w,\gamma) \,L^{y_w}, \qquad y_w = 1/\nu=1,
  \label{Upsdef}
\end{equation}
and taking the zero-temperature limit $T\sim u_t\to 0$.  Therefore,
keeping only the most relevant terms, the equilibrium
particle density in finite-size systems is expected to behave as
\begin{eqnarray}
  \varrho_e(w,\gamma,L) = \varrho_{\rm reg}(w,\gamma) +
    c_l\, u_w \ln L 
    + c_s L^{-y_n} {\cal D}_e(\Phi),\quad
  \label{fssrho}
\end{eqnarray}
where $\varrho_{\rm reg}(w,\gamma)$ is the same regular function
appearing in the infinite-volume scaling behavior~\cite{CPV-14,
  RV-21}, cf.~Eq.~\eqref{rhobeh}, and ${\cal D}_e$ a universal scaling
function (apart from a trivial multiplicative factor and normalization
of the argument).  Only the last term provides the genuine universal
scaling contribution of the critical modes (of course it depends on
the boundary conditions), while the other terms are not universal.  An
analogous behavior is found at classical transitions of 2D Ising
systems defined on finite-size square lattices, when considering the
finite-size behavior of the energy density (i.e., the derivative of
the free-energy density with respect to the
temperature)~\cite{CHPV-02}.

\begin{figure}[!t]
  \includegraphics*[scale=\graphicscale]{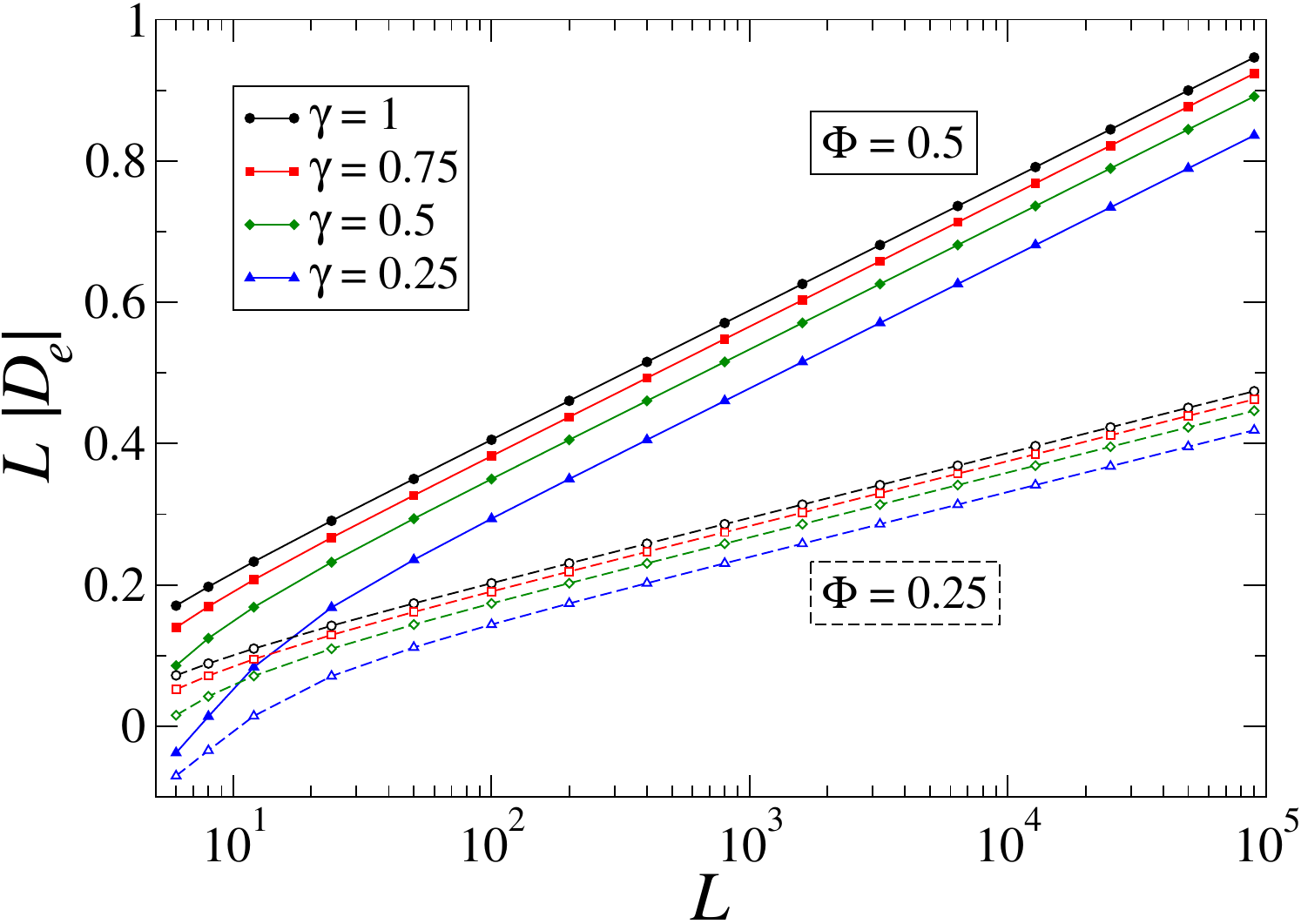}
  \caption{The absolute value of the equilibrium subtracted particle density $D_e$ as a
    function of the system size $L$, keeping the scaling variable $\Phi$
    fixed. Filled symbols denote numerical data for $\Phi=0.5$,
    while empty ones are for $\Phi=0.25$.  The various colors and
    symbols are for different values of the Hamiltonian parameter
    $\gamma$ (see legend).  Lines connecting symbols are drawn to
    guide the eye.  Note the logarithmic scale on the $x$-axis, so the
    observed straight-line behavior witnesses the expected large-$L$
    behavior $L D_e \sim \Phi \ln L$.}
  \label{figeqrho}
\end{figure}

We now focus on the subtracted particle density
\begin{eqnarray}
D_e(w,\gamma,L) \equiv \varrho_e(w,\gamma,L) - \varrho_c(\gamma),
\label{deltarhodeffss}
\end{eqnarray}
with $\varrho_c(\gamma)$ defined in Eq.~\eqref{rhoc}, whose
finite-size behavior can be easily derived using Eqs.~(\ref{fssrho})
and (\ref{rhoreg}).  Note that in the FSS limit, i.e., the
simultaneous limits $L\to\infty$ and $w\to 0$ keeping $\Phi = u_w
L^{y_w}$ fixed, the logarithmic term provides the leading
contribution. Therefore, when keeping $\Phi$ fixed, the finite-size
behavior~\eqref{fssrho} predicts the large-$L$ behavior
\begin{equation}
  L\, D_e(w,\gamma,L) \sim \Phi \ln L.
  \label{asyde}
\end{equation}
This is clearly confirmed by the numerical results shown in
Fig.~\ref{figeqrho}, for various choices of $\Phi$ and $\gamma$.

%%%%%%%%%%%%%%%%%%%%%%%%%%%%%%%%%%%%%%%%%%%%%%%%%%%%%%%%%%%%%%%%%%%%%%%%
%%%%%%%%%%%%%%%%%%%%%%%%%%%%%%%%%%%%%%%%%%%%%%%%%%%%%%%%%%%%%%%%%%%%%%%&

On the other hand, a standard asymptotic FSS is expected for the fermionic
correlation functions, such as
\begin{subequations}
  \label{pctf}
  \begin{eqnarray}
    G_P(x_1,x_2) & = & {\rm Tr}\left[ \rho_G \, (\hat c_{x_1}^\dagger 
      \hat c_{x_2}^\dagger + \hat c_{x_2} \hat c_{x_1}) \right], \\ 
    G_C(x_1,x_2) & = & {\rm Tr}\left[ \rho_G \, (\hat c_{x_1}^\dagger
      \hat c_{x_2} + \hat c_{x_2}^\dagger \hat c_{x_1}) \right].
  \end{eqnarray}
\end{subequations}
Indeed, assuming translational invariance, thus $G_\#(x)\equiv
G_\#(x_1,x_1+x)$, and in the zero-temperature limit, their equilibrium
FSS is given by~\cite{RV-21}
\begin{eqnarray}
 G_\#(x,w,L) \approx L^{-2 y_c} {\cal G}_{\#}(X,\Phi),\quad
 X\equiv x/L \neq 0
 \label{gcpscaeq}
\end{eqnarray}
($y_c=1/2$ is the RG dimension of the operator $\hat c_x$).

In passing, it is worth mentioning that one may also consider higher
derivatives of the free-energy density with respect to the chemical
potential.  However, their connection with the correlation functions
of the particle-density operator $\hat{n}_x$ is not
straightforward. In fact, due to the nontrivial derivative of the
Gibbs exponential operator~\cite{Wilcox-66}
\begin{equation}
  {d\over d\mu} e^{-\beta \hat{H}} = -\beta
  \int_0^1 e^{-z\beta\hat{H}} \hat{N}  e^{-(1-z)\beta \hat{H}} dz, 
  \label{derivmu}
\end{equation}
the derivative $\partial\varrho/\partial\mu$ is not related to the
connected expectation value $\langle \hat{N}^2 \rangle_c$.  Of course,
the relation~\eqref{rhobeh} can be easily derived using the fact that
\begin{equation}
  {\rm Tr} \left[ {d\over d\mu} e^{-\beta \hat{H}}\right]
  = -\beta \, {\rm Tr} \big[ e^{-\beta\hat{H}} \hat{N} \big].
  \label{trder}
\end{equation}

In conclusion, the above scaling analyses show that the asymptotic
behavior of the particle density at the critical point is
characterized by competing contributions from different sources: the
regular background term of the free-energy density, a peculiar
logarithmic resonance term, and the critical scaling term controlled
by the critical exponents. Even the subtracted particle
density~\eqref{deltarhodeffss} does not show an asymptotic scaling
behavior, which turns out to be hidden by a logarithmic contribution
and the regular term arising from mixings with the identity
operator. In the following we investigate whether, and how,
out-of-equilibrium conditions arising from quantum quenches can
disentangle the various contributions, recovering a well defined
universal out-of-equilibrium FSS behavior.

\section{Dynamics after a quench}
\label{outeq}

The dynamics of quantum many-body systems is often studied by
considering protocols based on instantaneous quantum quenches of the
model parameters (see, e.g., Refs.~\onlinecite{Niemeijer-67, BM-70,
  CC-06, PEF-12, GS-12, Trotsky-etal-12, Gring-etal-12, CC-16,
  Cardy-16, BRI-16, PRV-18, RV-21, TV-23}), or protocols entailing
slow changes of such parameters like those associated with the
so-called Kibble-Zurek (KZ) problem (see, e.g.,
Refs.~\onlinecite{Kibble-80, Zurek-85, Zurek-96, PSSV-11, CEGS-12,
  Biroli-16, RV-20-kz, RV-21, TV-22, DV-23}).

In this section we focus on the quantum evolution arising from a
so-called soft quench around the critical point, for which the
variation of the parameters associated with the quench is sufficiently
small to maintain the system close to criticality.  We address the
out-of-equilibrium scaling behavior along the post-quench critical
quantum evolution, paying particular attention to the behavior of the
particle density.

We also report numerical results for the out-of-equilibrium evolution
after the quench, up to very large $O(10^4)$ lattice sizes,
obtained by a straightforward diagonalization of the
Hamiltonian~\eqref{kitaev} [see also Eq.~\eqref{H-harmonic}].
With antiperiodic boundary conditions, this can be done by decoupling
$\hat H$ into a sum of $L/2$ independent terms, each of them acting in
the four-dimensional Hilbert subspace generated by the $k$ and $-k$
modes, for a given value of $n$ in Eq.~\eqref{kval}, and then
exploiting the conservation of fermion parity to further reduce it to
two dimensions.  Then, the unitary time evolution can be easily
computed numerically by a direct integration of the Schr\"odinger's
equation on each of such subspaces.

\subsection{Soft quench protocol}
\label{quprot}

We perform an instantaneous quench of the chemical-potential
parameter, from $\mu\neq \mu_c$ to $\mu_c$, or correspondingly from
$w\equiv (\mu_c-\mu)/2\neq 0$ to $w_c=0$, in such a way to study the
critical out-of-equilibrium quantum evolution.  In practice, we
consider the following protocol: (i) At $t=0$ the system is prepared
in the ground state $|\Psi_{\rm GS}(w)\rangle$ of the Hamiltonian
$\hat H(w,\gamma)$, cf. Eq.~\eqref{kitaev}, for a given value of
$w\neq 0$. (ii) At $t>0$, the system evolves unitarily driven by the
critical Hamiltonian $\hat H(\mu_c,\gamma)$, i.e.
\begin{equation}
  i {d\over dt} |\Psi(t)\rangle = \hat H(w_c,\gamma)\, |\Psi(t)\rangle,
  \quad |\Psi(0)\rangle \equiv |\Psi_{\rm GS}(w) \rangle. 
  \label{scheq}
\end{equation}
We remark that we only consider soft quenches starting from initial
conditions close to the critical point (i.e., for small values of
$|w|$), so that the system stays always within the critical regime
during the post-quench quantum evolution.

The arising out-of-equilibrium dynamics can be monitored using the
particle density and the fermionic correlation functions, analogous to
the equilibrium definitions in Eqs.~\eqref{rhobeh} and~\eqref{pctf},
respectively, replacing the Gibbs density matrix with the
time-dependent density matrix of the evolving (pure) state
\begin{equation}
  \rho_\Psi(t) = |\Psi(t)\rangle \langle \Psi(t)|.
  \label{rhopsit}
\end{equation}

\subsection{Out-of-equilibrium finite-size scaling}
\label{dynsca}

As shown in Refs.~\onlinecite{PRV-18, RV-21, TV-22}, the post-quench
quantum evolution of standard observables characterized by an asymptotic
equilibrium FSS, such as the fermionic correlations of the fermionic
Kitaev wires and the longitudinal spin correlations of the quantum
$XY$ chain, develops an out-of-equilibrium FSS behavior at quantum
transitions. This is essentially obtained by adding a further
dependence on the time scaling variable $\Theta \sim t/\tau $ to the
equilibrium FSS behaviors, where $\tau$ is the time scale of the
critical modes, which is expected to be related to the gap
$\Delta_c(L)$ (energy difference of the lowest eigenmodes) at the
critical point, i.e., $\tau\sim 1/ \Delta_c(L) \sim L^{z}$.  Actually,
since the gap corresponds to the lowest excitation energy (for $k =
\pm \pi/L$) and thus its finite-size dependence at the critical point
is given by
\begin{equation}
  \Delta_c(L,\gamma) = E\left(k=\frac{\pi}{L},\mu=\mu_c,\gamma\right)
  = \frac{2\pi\gamma}{L} + O(L^{-3}),
\label{deltabeh}
\end{equation}
where $E(k,\mu,\gamma)$ is the function defined in Eq.~\eqref{Ek-XY},
we define
\begin{equation}
  \Theta = {\gamma \, t\over L} \sim \Delta_c(L,\gamma)\,t,
  \label{thetadef}
\end{equation}
including a $\gamma$-dependent normalization.

\begin{figure}[!t]
  \includegraphics*[scale=\graphicscale]{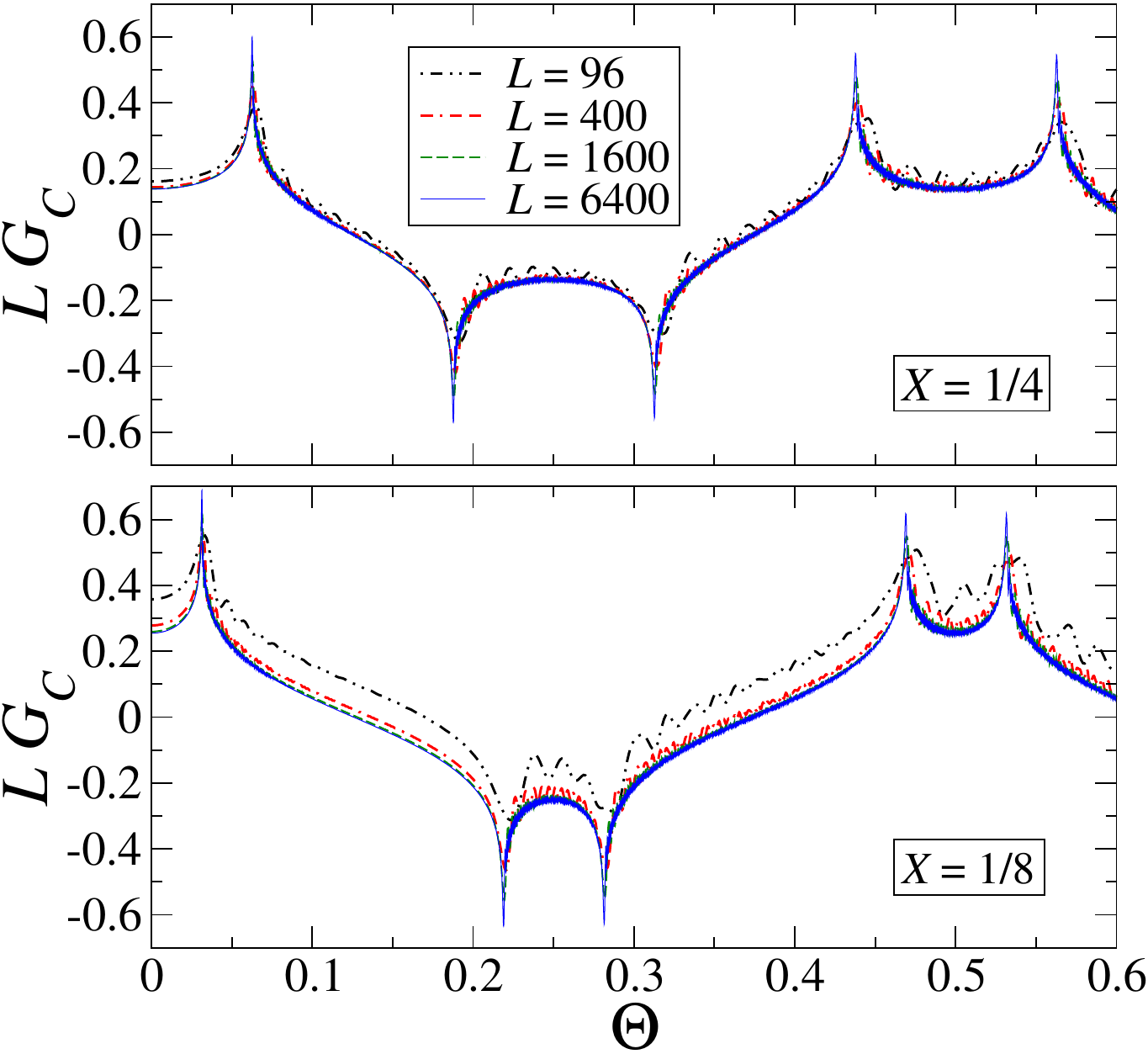}
  \caption{Out-of-equilibrium FSS of the fermionic correlation $G_C$
    as a function of the rescaled time $\Theta$, for fixed $X=x/L=1/4$
    (upper panel) and $X=1/8$ (lower panel), after a soft quench from
    the initial condition $\Phi \equiv u_w L = -0.5$ to the critical
    point $\Phi=0$, for $\gamma=1$ [cf.~Eq.~\eqref{gxscaout}].  The
    various curves correspond to different system sizes $L$, as
    indicated in the legend. The approach to the asymptotic
    out-of-equilibrium FSS is consistent with a simple $1/L$ power-law
    behavior, as expected~\cite{RV-21}.}
  \label{gcouteq}
\end{figure}

Then, along the quench protocol outlined in Sec.~\ref{quprot}, the
fixed-time two-point functions $G_P(t,x,w,L)$ and $G_C(t,x,w,L)$ are
expected to develop the out-of-equilibrium FSS~\cite{PRV-18, RV-21}
\begin{equation}
  G_\#(t,x,w,L) \approx L^{-2y_c} \,{\cal
    G}_\#(X,\Phi,\Theta), \quad y_c = 1/2,
  \label{gxscaout}
\end{equation}
asymptotically in the out-of-equilibrium FSS limit
(i.e., the large-$L$ and large-$t$ limits, keeping the scaling
variables $X\equiv x/L$, $\Phi \equiv u_w L^{y_w}$, and $\Theta$
fixed).  Note that the dependence on $\Phi$ is essentially related to
the initial condition, while no scaling variable is associated with
the post-quench value of $w=w_f$, because it is always set to zero
(otherwise the additional scaling variable $\Phi_f=u_{w_f} L^{y_w}$
should have been added).  The out-of-equilibrium FSS functions ${\cal
  G}_\#$ are universal, therefore they must be independent of
$\gamma$, apart from a multiplicative factor and possible nonuniversal
normalizations of the scaling variables $\Phi$ and $\Theta$.
Actually, their definitions~\eqref{Upsdef} and~\eqref{thetadef}
already contain the correct $\gamma$ dependence to avoid further
$\gamma$-dependent normalizations.  The asymptotic out-of-equilibrium
FSS is expected to be approached with power-law scaling corrections,
like at equilibrium.
Figure~\ref{gcouteq} shows some numerical results for $G_C$, clearly
confirming the out-of-equilibrium FSS in Eq.~\eqref{gxscaout}. We also
note the presence of spikes at finite values of the rescaled time $\Theta$,
which will be discussed in detail later.

%%%%%%%%%%%%%%%%%%%%%%%%%%%%%%%%%%%%%%%%%%%%%%%%%%%%%%%%%%%%%%%%%%%%%%%%
%%%%%%%%%%%%%%%%%%%%%%%%%%%%%%%%%%%%%%%%%%%%%%%%%%%%%%%%%%%%%%%%%%%%%%%%

The above out-of-equilibrium FSS behaviors have been obtained by a
natural extension of their equilibrium scaling behaviors, simply
adding a time dependence through the time scaling variable
$\Theta$. However, it is not straightforward to extend this simple
picture to the out-of-equilibrium time dependence of quantities whose
equilibrium behavior is affected by regular and singular contributions
involving the identity operator, such as those appearing in the
particle density behavior at the critical point.  As discussed in
Sec.~\ref{sec:scaling}, the scaling term of the equilibrium particle
density $\varrho_e$, and also of its subtracted definition $D_e$, is
generally hidden by the analytical background and the logarithmic
resonance term, cf.~Eq.~\eqref{fssrho}.

The question is whether the modes related to the identity operator
share the same time scale $\tau\sim L^z$ of the critical modes, or
their time scale $\tau_I$ is significantly shorter, as may be
suggested by the fact they are expected to arise from short-range
quantum fluctuations. If the ratio $\tau_I/\tau$ vanishes in the
large-$L$ limit, then the out-of-equilibrium particle density may
develop a scaling behavior characterized by the same power law of the
scaling term of the equilibrium particle density,
cf. Eq.~\eqref{rhosca}. An analogous phenomenon occurs along the
post-quench critical relaxational flow at classical thermal
transitions~\cite{PV-24}, where the energy density shows an
out-of-equilibrium FSS behavior, even though it does not scale at
equilibrium, like the equilibrium particle density at quantum
transitions driven by the chemical potential.

To investigate the out-of-equilibrium behavior of the particle density
under the post-quench critical quantum evolution, we consider the
subtracted particle density
\begin{equation}
  D(t,w,\gamma,L) \equiv {1\over L} {\rm Tr}\,[\rho_\Psi(t) \, \hat{N}]
  - \varrho_c(\gamma),
  \label{diffet}
\end{equation}
where $\varrho_c$ is the $\gamma$-dependent value of the the particle
density at the critical point in the thermodynamic limit,
cf. Eq.~\eqref{rhoc}.  If the time scales of the different
contributions differ substantially, then the post-quench quantum
evolution may disentangle the scaling contribution from the terms
arising from the mixing with the identity operator.  As we shall see,
the post-quench out-of-equilibrium behavior of the particle density
turns out to develop the nontrivial asymptotic out-of-equilibrium FSS
\begin{equation}
  D(t,w,\gamma,L) \approx L^{-y_n} {\cal D}(\Phi,\Theta),
  \label{dtscal}
\end{equation}
which is analogous to a standard scaling behavior, like the case of
the fermionic correlations, cf. Eq.~\eqref{gxscaout}.

However, like the energy density at classical transitions with
negative specific-heat exponent~\cite{PV-24}, the out-of-equilibrium
FSS function ${\cal D}$ develops a nontrivial singularity for
$\Theta\to 0$, essentially related to the fact that at $\Theta=0$, and
therefore at equilibrium, the subtracted particle density does not
show a universal asymptotic FSS, behaving as $L^{-1} \ln L$. Indeed,
by matching the out-of-equilibrium FSS of the subtracted particle
density, put forward in Eq.~\eqref{dtscal}, with the leading
logarithmic term of the equilibrium behavior [cf.~Eq.~\eqref{asyde}],
one would predict a logarithmic divergence of the scaling function
${\cal D}(\Phi,\Theta)$ when $\Theta\to 0$ keeping $\Phi$ fixed.

\subsection{Numerical results}
\label{numres}

%%%%%%%%%%%%%%%%%%%%%%%%%%%%%%%%%%%%%%%%%%%%%%%%%%%%%%%%%%%%%%%%%%%%%%%%
\begin{figure}[!t]
  \includegraphics*[scale=\graphicscale]{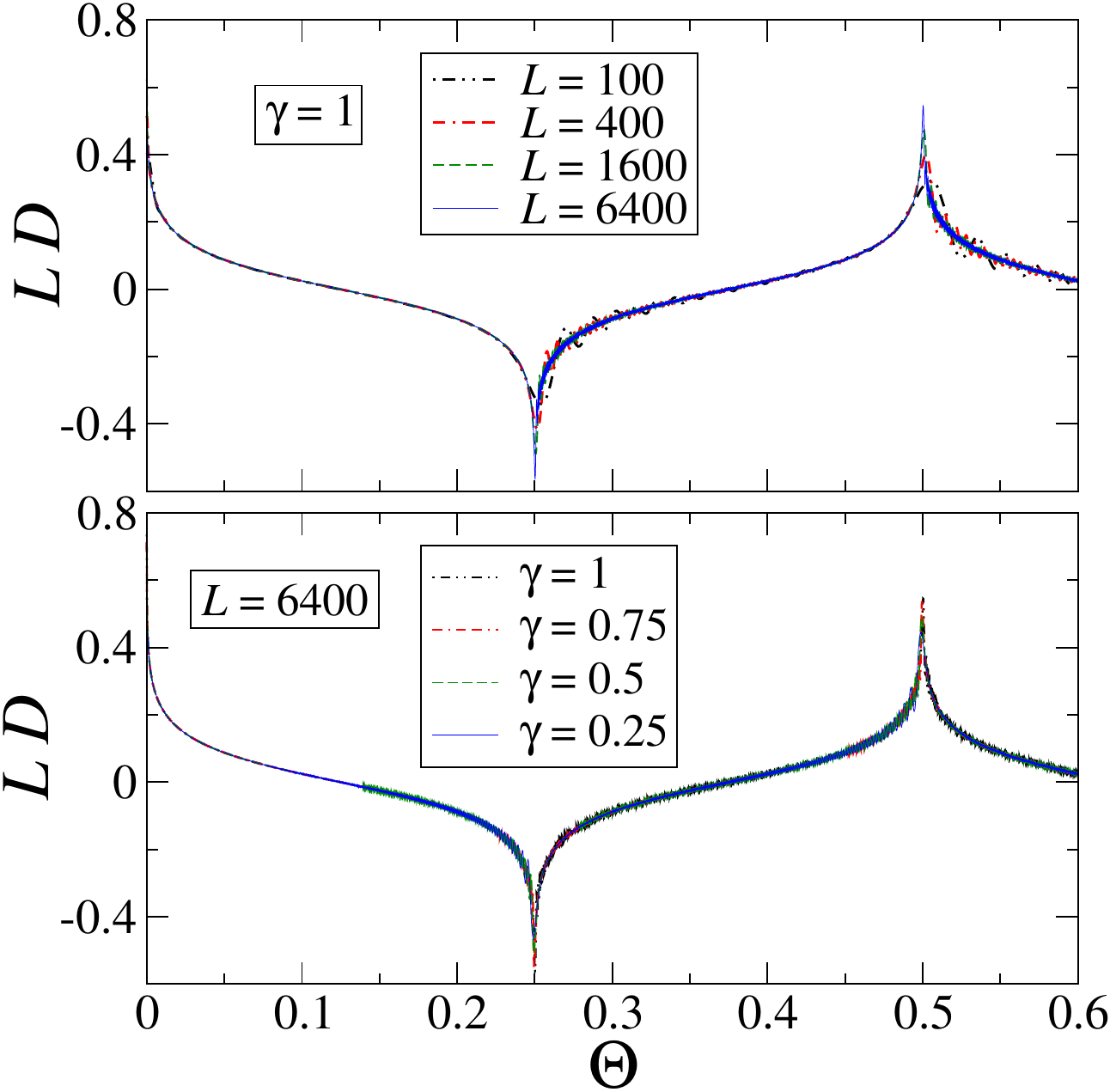}
  \caption{Scaling of the subtracted particle density $D$ with time,
    after a quench from $\Phi = -0.5$ to the critical point
    [cf.~Eq.~\eqref{dtscal}].  Upper panel: curves are for different
    values of $L$, while $\gamma=1$ is kept fixed.  Lower panel:
    curves are for different values of $\gamma$, while $L=6400$ is
    kept fixed: a collapse of the various curves with $\gamma$
    supports universality of the scaling function ${\cal
      D}(\Phi,\Theta)$.}
  \label{densitytheta}
\end{figure}
%%%%%%%%%%%%%%%%%%%%%%%%%%%%%%%%%%%%%%%%%%%%%%%%%%%%%%%%%%%%%%%%%%%%%%%%

The typical out-of-equilibrium behavior of the subtracted particle
density $D(t,w,\gamma,L)$ after a soft quench is reported in
Fig.~\ref{densitytheta}, for initial conditions corresponding to a
fixed scaling variable $\Phi=-0.5$ [cf.~Eq.~\eqref{Upsdef}].  Analogous
results are obtained for other values of $\Phi$, both negative
and positive (not shown). The
numerical data for different sizes $L$ (upper panel) nicely support
the scaling ansatz~\eqref{dtscal}, at generic values of $\Theta$.  We
have also performed simulations for various values of $\gamma$ (lower
panel), showing that the scaling function ${\cal D}(\Phi,\Theta)$ is
universal (i.e., independent of $\gamma$).  Aside from this expected
behavior, two special occurrences which correspond to specific values
of $\Theta$ have to be carefully addressed.

\begin{figure}[!t]
  \includegraphics*[scale=\graphicscale]{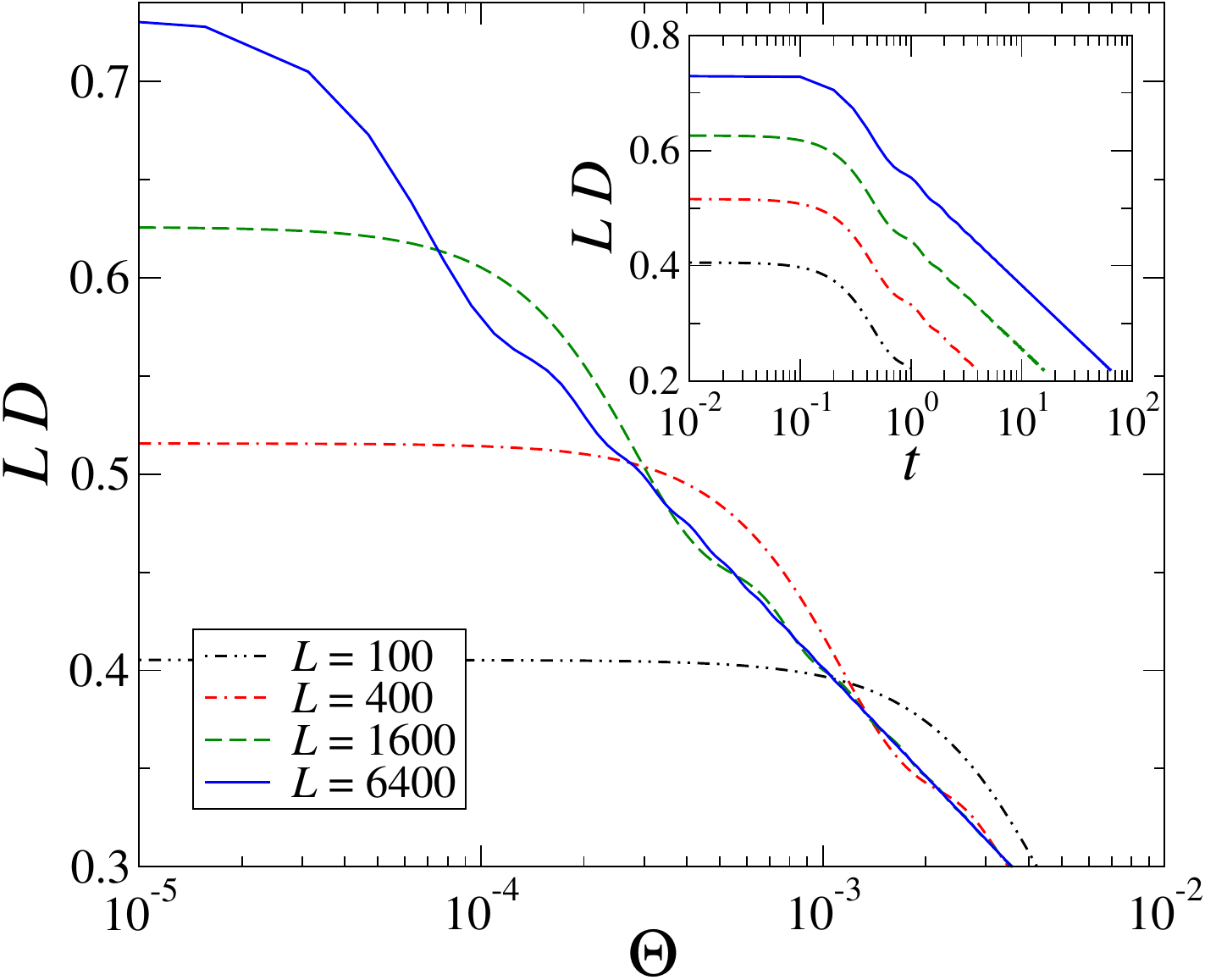}
  \caption{A zoom of Fig.~\ref{densitytheta} upper panel, for small
    values of the rescaled time ($\Theta \leq 10^{-2}$).  The $x$-axis
    is in log scale, such that a straight-line behavior denotes a
    logarithmic divergence of the scaling function $L D\approx {\cal
      D}(\Phi,\Theta)$ for $\Theta \to 0$, as ${\cal D} \sim \log
    \Theta$.  The inset shows the same data, but with the real time
    $t$ (and not the rescaled time $\Theta$) on the $x$-axis.  Note
    that curves start deviating from the equilibrium value at a time
    $\tau_s \sim O(L^0)$ which is independent of $L$.}
  \label{theta0lim}
\end{figure}

First of all, at small values of $\Theta$, our numerics clearly
evidences a logarithmic divergence with $L$ of the product $L \,
D(t,w,\gamma,L)$, which is somehow reconstructing the logarithmic
equilibrium behavior arising from the resonance between the energy and
identity operators, as already mentioned at the end of
Sec.~\ref{dynsca}. This is shown in Fig.~\ref{theta0lim}, displaying a
zoom of the curves in the upper panel of Fig.~\ref{densitytheta}, for
$\Theta \to 0$.  We also note that the out-of-equilibrium FSS sets in
for $\Theta \gtrsim \tau_s/L$, where the time scale $\tau_s$ turns out
to be independent of $L$ (see also the inset, which displays the same
data as a function of the bare time $t$).  Thus, $\tau_s$ is the time
scale after which the logarithmic singularity in $\Theta$, for $\Theta
\to 0$, starts emerging.  Such time should be identified, or at least
strictly connected, with the time scale $\tau_I$ of the modes related
to the identity operator, which is needed to equilibrate short-range
fluctuations, i.e., $\tau_s \sim \tau_I \sim O(L^0)$ with increasing
$L$.

%%%%%%%%%%%%%%%%%%%%%%%%%%%%%%%%%%%%%%%%%%%%%%%%%%%%%%%%%%%%%%%%%%%%%%%%
%%%%%%%%%%%%%%%%%%%%%%%%%%%%%%%%%%%%%%%%%%%%%%%%%%%%%%%%%%%%%%%%%%%%%%%%

\begin{figure}[!t]
  \includegraphics*[scale=\graphicscale]{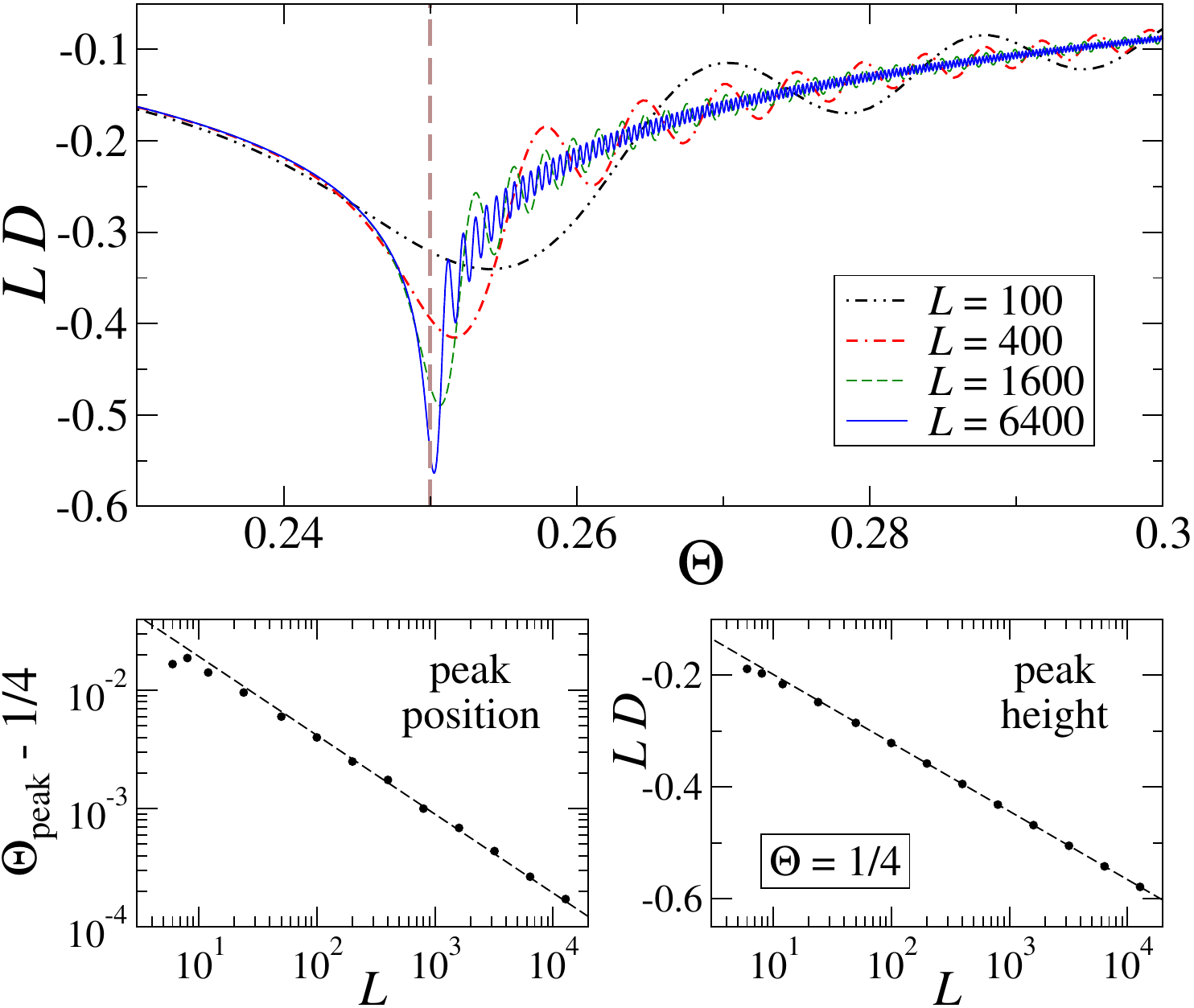}
  \caption{A zoom of Fig.~\ref{densitytheta} upper panel, around the
    first spike that develops for $\Theta \to \Theta_1 = 1/4$
    (vertical dashed line).  Bottom left panel: the location of the
    peak with $L$; the asymptotic value $1/4$ is approached with
    $O(L^{-2/3})$ corrections~\cite{RV-20} (represented by the dashed
    straight line to guide the eye, note the loglog scale of the
    plot).  Bottom right panel: the behavior of $D$ with $L$ at
    $\Theta_1$; the straight line is a logarithmic fit $L D = a + b
    \log L$ for $L > 10^3$, with $a$ and $b$ fitting parameters (note
    the log scale on the $x$-axis).}
  \label{spikefig}
\end{figure}

Second, it is worth stressing that the curves reported in
Fig.~\ref{densitytheta} for the subtracted particle density develop
peculiar spikes at finite values of $\Theta$.  For example, the spikes
occurring around $\Theta=0.25$ are highlighted in Fig.~\ref{spikefig},
showing a strong evidence of the presence of another logarithmic
divergence of $L \, D$, evaluated at $\Theta=0.25$, with $L$ (bottom
right panel).  Analogous results are found for $\Theta=0.5$ and, in
general, for multiples of $\Theta=0.25$ (not shown).  These
singularities can be related to revival phenomena (see, e.g.,
Refs.~\onlinecite{IR-11, RI-11, BRI-12, HHH-12, KLM-14, Cardy-14, JH-17,
  MAC-20, RV-20}), due to the finite size of the system.  Indeed, they
appear at times
\begin{equation}
  \Theta_k \equiv {\gamma t_{k}\over L},\qquad
  t_k = {k L \over 2v_m}\,,\qquad v_m = {2\gamma},
  \label{tk}
\end{equation}
for $k = 1,2, \ldots$, where $v_m$ is the maximum velocity of the
quasi-particle modes at the critical point~\cite{LR-72, CC-05,
  CEF-12}.  In fact, our numerics shows that the first emerging spikes
are asymptotically (i.e., for $L\to\infty$) located at $\Theta_{1} =
1/4$ and $\Theta_{2} = 1/2$ with a great accuracy (see
Fig.~\ref{densitytheta}).  Such values of $\Theta$ correspond to the
rescaled time for the quasi-particle modes to run across half lattice
size (or multiple of it), thus confirming their interpretation in
terms of revival phenomena.  Finite-size corrections to the
$\Theta$-location of such peaks are power-law suppressed with $L$, as
reported in the bottom left panel of Fig.~\ref{spikefig} and already
found in Ref.~\onlinecite{RV-20} for similar revival phenomena.

%%%%%%%%%%%%%%%%%%%%%%%%%%%%%%%%%%%%%%%%%%%%%%%%%%%%%%%%%%%%%%%%%%%%%%%%
%%%%%%%%%%%%%%%%%%%%%%%%%%%%%%%%%%%%%%%%%%%%%%%%%%%%%%%%%%%%%%%%%%%%%%%%

\begin{figure}[!t]
  \includegraphics*[scale=\graphicscale]{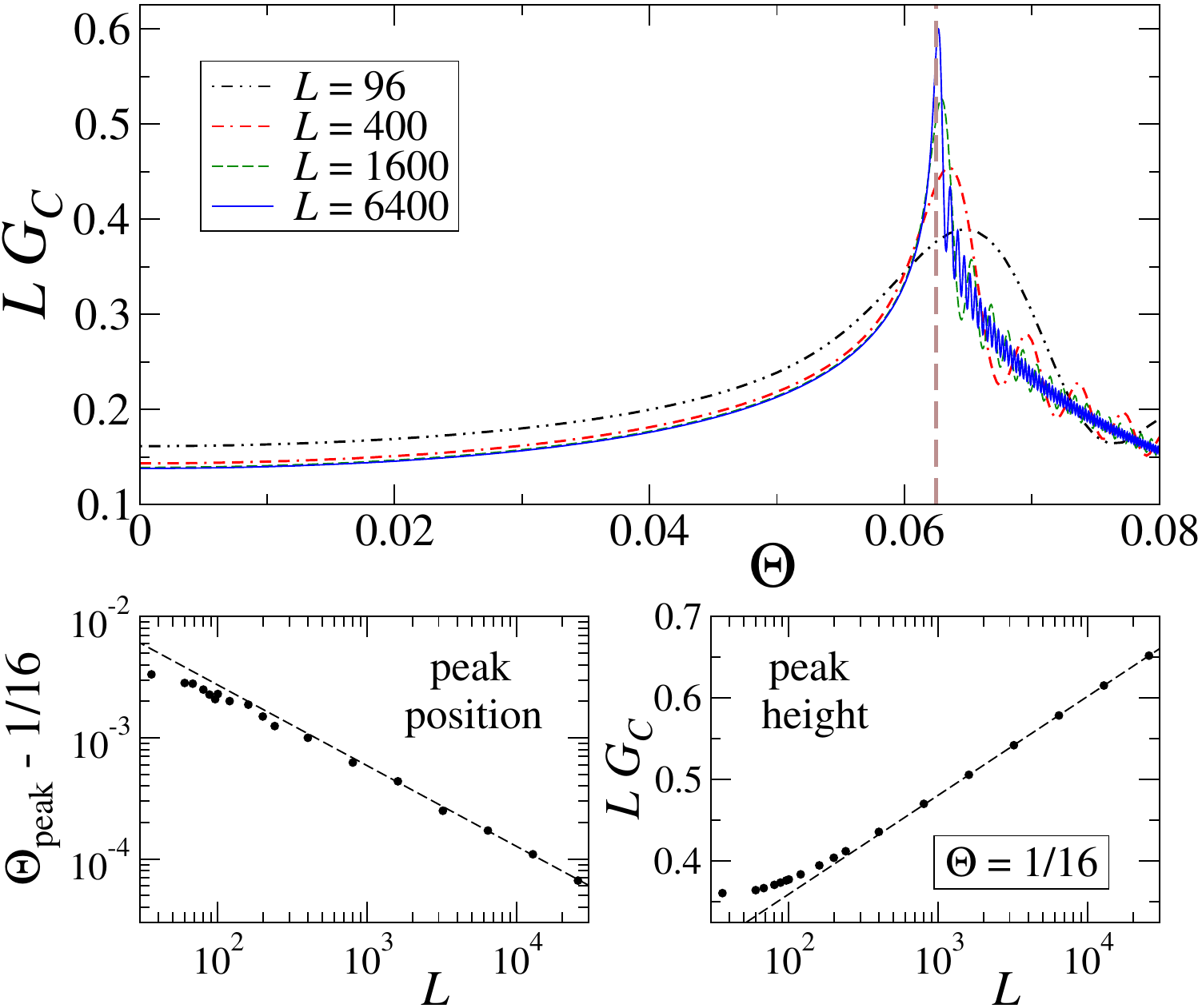}
  \caption{A zoom of Fig.~\ref{gcouteq} upper panel, for $\Theta \in
    [0, 0.08]$, which highlights the spike that develops for $\Theta
    \to \Theta_1 \equiv 0.0625$ (vertical brown dashed line).  Also
    note that, for $\Theta \to 0$, curves approach the expected
    equilibrium scaling behavior in Eq.~\eqref{gcpscaeq}.  Bottom left
    panel: the location of the peak with $L$; analogously as for the
    subtracted particle density, finite-size deviations from the
    asymptotic value $1/16$ scale as $L^{-2/3}$ (dashed line), see
    also Ref.~\onlinecite{RV-20}, where analogous revival phenomena
    are reported.  Bottom right panel: the behavior of $G_C$ with $L$,
    evaluated at $\Theta_1$; the straight line is a logarithmic fit $L
    G_C = a + b \log L$ of the numerical data for $L > 10^3$, with $a$
    and $b$ fitting parameters.}
  \label{gcspikes}
\end{figure}

It is also worth pointing out that analogous spikes can be observed in
the post-quench behavior of the fermionic correlations see
Fig.~\ref{gcouteq}.  Again, they appear to be associated with
logarithmic divergences in $L$ (which have been previously
overlooked~\cite{NRV-19}), as reported in Fig.~\ref{gcspikes} for the
peak in the correlation function $G_C$ with $X=1/4$ emerging at
$\Theta=0.0625$ (the upper panel is a magnification of the data in
Fig.~\ref{gcspikes} around $\Theta=0.0625$, while the bottom right
panel displays the behavior of $L \, G_C$, evaluated exactly at
$\Theta=0.25$, with $L$).  Their location differs from those of the
particle density, essentially because they are nonlocal observables
characterized by a scaling distance $X$. Actually it corresponds to
the rescaled time to run across half rescaled distance $X$, and the
complementary value $1-X$, at the maximum speed of the quasi-particle
modes, i.e., the first logarithmic spikes are (asymptotically) located
at
\begin{equation}
  \Theta_1 = {\gamma X\over 2 v_m},\qquad
  \Theta_2 = {\gamma(1-X)\over 2 v_m},
  \label{spikelocs}
\end{equation}
respectively, see Fig.~\ref{gcouteq}. This may be interpreted as an
emerging singularity analogous ot that the particle density, when the
quasi-particle modes starting at relative distance $X$ meet.
Analogously as before, at finite size, the position of the peaks
approaches the asymptotic value as a power-law with $L$
(bottom right panel of Fig.~\ref{gcspikes}).

%%%%%%%%%%%%%%%%%%%%%%%%%%%%%%%%%%%%%%%%%%%%%%%%%%%%%%%%%%%%%%%%%%%%%%%%
%%%%%%%%%%%%%%%%%%%%%%%%%%%%%%%%%%%%%%%%%%%%%%%%%%%%%%%%%%%%%%%%%%%%%%%%

The above results fully confirm the out-of-equibrium FSS behavior
after a soft quench, put forward in Sec.~\ref{dynsca}, and in
particular that of the particle density given in
Eq.~\eqref{dtscal}. Since the particle density should be quite
accessible experimentally and numerically, its out-of-equilibrium
behavior under quantum quenches may provide a further effective probe
of the universal features at quantum transitions, unlike its
equilibrium behavior that is essentially dominated by nonuniversal
short-range fluctuations.

\subsection{Out-of-equilibrium scaling in the thermodynamic limit}
\label{outthlim}

The previous results show that the subtracted particle density
develops a peculiar out-of-equilibrium FSS, according to
Eq.~\eqref{dtscal}, characterized by various singularities of the
corresponding scaling function, in particular in the $\Theta\to 0$
limit keeping $\Phi$ fixed. We now focus on the out-of-equilibrium
scaling behavior of the subtracted particle density in the
thermodynamic limit.

To derive a scaling ansatz from the FSS behavior~\eqref{dtscal}, we
note that the thermodynamic limit corresponds to the limit
$L/\xi\to\infty$, where $\xi\sim w^{-\nu}$ is the correlation length of
the system. This is achieved by taking the $\Phi\to\infty$ limit
keeping the product $\Phi^{z\nu} \Theta = \Phi \Theta = \gamma
\,u_w\,t$ fixed. Actually, for simplicity, we consider the scaling
variable
\begin{equation}
  \theta = w \, t,
\end{equation}
so that $\theta\approx \Phi\Theta$ apart from irrelevant subleading
terms in the scaling limit [keeping into account that $u_w=
w/\gamma + O(w^2)$].  Then, the thermodynamic limit of the
out-equilibrium FSS Eq.~\eqref{dtscal} can be straightforwardly
obtained by taking the limit $\Phi\to \infty$ keeping $\theta$ fixed,
so that
\begin{equation}
  D_\infty(t,w,\gamma)= D(t,w,\gamma,L\to\infty) \approx {w\over
    \gamma} \, {\cal D}_\infty(\theta).
  \label{dtscalinf}
\end{equation}
Note that we keep the $\gamma$ dependence in the prefactor of the
r.h.s.~of Eq.~\eqref{dtscalinf}, which turns out to be useful to check
the universality of the scaling behavior with respect to the
Hamiltonian parameter $\gamma$.

The above out-of-equilibrium scaling behavior can be checked using the
known analytical results for the post-quench time dependence of the
transverse magnetization of the quantum $XY$ chain in the
thermodynamic limit, already reported in
Refs.~\onlinecite{Niemeijer-67,BM-70}, where the post-quench
time dependence of the transverse magnetization is expressed in terms
of an integral over the momenta (as a function of the coupling
$g_i>g_c$ of the initial ground state before the quench, the post-quench
coupling $g$, and the parameter $\gamma$). In the thermodynamic limit,
the boundary conditions become irrelevant, therefore these
computations apply also to the particle density of the Kitaev wire in
the corresponding infinite-size limit.

\begin{figure}[!t]
  \includegraphics*[scale=\graphicscale]{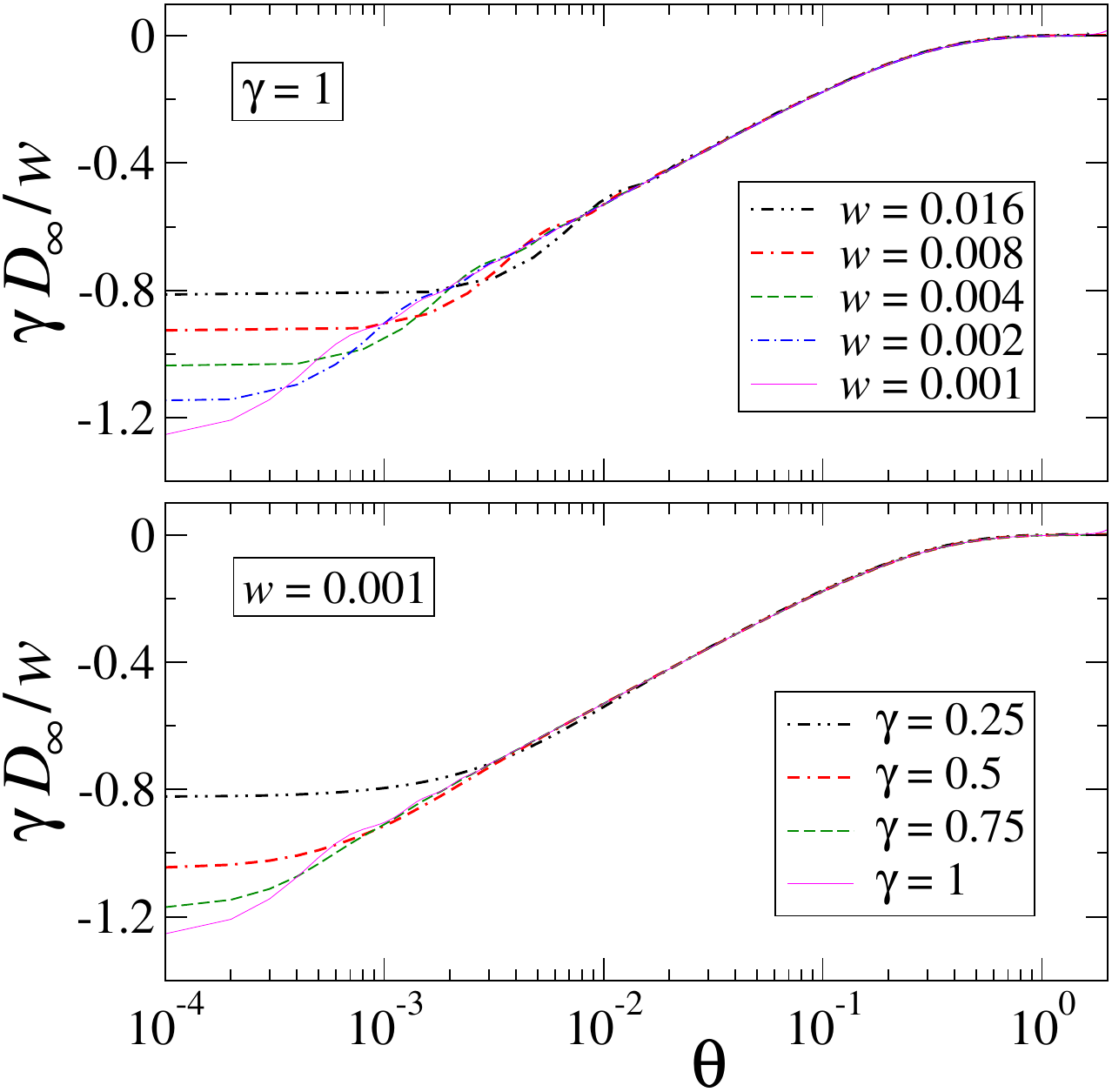}
  \caption{Scaling of the subtracted particle density $D_\infty$ with
    time, in the thermodynamic limit [cf.~Eq.~\eqref{dtscalinf}].
    Upper panel: curves are for different
    values of $w$, keeping $\gamma=1$ fixed. Lower panel: curves are for
    different values of $\gamma$, keeping $w=10^{-3}$ fixed.}
  \label{infsizelim}
\end{figure}

Applying the analytical expressions of
Refs.~\onlinecite{Niemeijer-67,BM-70} to the soft quench protocol
outlined in Sec.~\ref{quprot}, and taking the out-of-equilibrium
scaling limit $w = g_i-g_c\to 0$ and $t\to\infty$ keeping $\theta$
fixed, we obtain the curves shown in Fig.~\ref{infsizelim}. They
clearly show the scaling behavior~\eqref{dtscalinf} (upper panel),
and universality with respect to variations of $\gamma$ (lower panel).
Moreover, the scaling curve displays again a logarithmic divergence
for $\theta\to 0$, analogously to what has been observed
in Fig.~\ref{theta0lim}, in the FSS framework.

%%%%%%%%%%%%%%%%%%%%%%%%%%%%%%%%%%%%%%%%%%%%%%%%%%%%%%%%%%%%%%%%%%%%%%%%
%%%%%%%%%%%%%%%%%%%%%%%%%%%%%%%%%%%%%%%%%%%%%%%%%%%%%%%%%%%%%%%%%%%%%%%%

\section{Quasi-adiabatic Kibble-Zurek protocol}
\label{KZprot}

We now discuss the out-of-equilibrium behavior of the particle density
arising from slow changes of the chemical potential approaching the
quantum transition, such as the dynamic protocols related to the
so-called KZ problem, related to the defect production when crossing
continuous transitions from disordered to ordered phases (see, e.g.,
Refs.~\onlinecite{Kibble-80, Zurek-85, Zurek-96, PSSV-11, CEGS-12,
  Biroli-16, Dziarmaga-10, RV-21}).

Quasi-adiabatic evolutions at quantum transitions can be obtained by
slowly varying the Hamiltonian parameter $\mu$, according to the
linear time dependence
  \begin{equation}
    \mu(t) = \mu_c - 2 w(t),\qquad  w(t) = -t/t_s,
    \label{wtkz}
  \end{equation}
with a large time scale $t_s$.  More precisely, we consider the
following quasi-adiabatic protocol: (i) The quantum evolution of
finite fermionic wires of size $L$ starts at a time $t_i<0$ from the
ground state $|\Psi_{\rm GS}(w_i)\rangle$ associated with the initial
value $w_i= -t_i/t_s > 0$.  (ii) Then the system evolves unitarily
according to the Schr\"odinger equation
\begin{equation}
  i {d\over dt} |\Psi(t)\rangle = \hat H[w(t),\gamma] \, |\Psi(t)\rangle,
  \quad |\Psi(0)\rangle = |\Psi_{\rm GS}(w_i) \rangle, 
  \label{unitdyn}
\end{equation}
where $w$ varies linearly as in Eq.~\eqref{wtkz} up to the final
value $w_f=0$, corresponding to $t_f=0$, thus $w(t)\ge 0$ along the
whole protocol.  If we assume $w_i$ fixed with increasing $t_s$, then
$t_i \to -\infty$ in the large $t_s$ limit.
  
At a quantum transition, the development of an out-of-equilibrium
dynamics is inevitable (in the thermodynamic limit $L\to\infty$,
before taking the critical limit) even for very slow changes of the
parameter $w$, because large-scale modes are unable to equilibrate the
long-distance critical correlations emerging at the transition
point. As a consequence, when starting from equilibrium states at the
initial value $w_i$, the system cannot pass through equilibrium states
associated with the values of $w(t)$ at the transition point, thus
departing from an adiabatic dynamics before arriving at $w=0$. Such a
departure develops peculiar out-of-equilibrium scaling phenomena in
the limit of large time scale $t_s$ of the time variation of
$w(t)$. In particular, during the quantum evolution of finite systems
under KZ protocols, an out-of-equilibrium FSS emerges in the large-$L$
and large-$t_s$ limits, keeping the appropriate scaling variables
fixed (see, e.g., Refs.~\onlinecite{RV-21, RV-20-kz, DV-23, TV-23}).

The results of this kind of dynamics can be monitored by looking at
some observables and correlations at fixed time, such as the fermionic
correlations $G_P$ and $G_C$ [cf.~Eqs.~\eqref{pctf}], replacing
$\rho_G$ with $\rho(t)=|\Psi(t)\rangle \langle \Psi(t)|$.  Their
out-of-equilibrium FSS along the KZ protocol can be written in terms
the scaling variables~\cite{RV-21}
\begin{equation}
  S_1 \propto w(t) L^{y_w},\qquad  S_2 \propto t\, \Delta_c(L) \sim t/L^z,
  \label{s12def}
\end{equation}
where $\Delta_c$ is the critical gap, or more convenient combinations
such as
\begin{equation}
  \Upsilon \propto -\frac{S_2}{S_1} \propto \frac{t_s}{L^{\zeta}}, \qquad
  \Omega \propto \frac{S_2}{\Upsilon^{\kappa}} \propto \frac{t}{t_s^{\kappa}},
  \label{scalvars} 
\end{equation}
where
\begin{equation}
\zeta = y_w + z = 2, \qquad \kappa={z\over y_w + z}={1\over 2}.
\label{kappadef}
\end{equation}
Actually, analogously to the definitions of the scaling variables
$\Phi$ and $\Theta$ associated with the quench protocol
[cf.~Eqs.~\eqref{Upsdef} and~\eqref{thetadef}, respectively], we may
allow for a $\gamma$-dependent normalization of the scaling variables,
to simplify the universality checks with respect to variations of
$\gamma$, without including further factors in the dependence of the
scaling variables. Taking into account the fact that $u_w \sim
w/\gamma$ [cf.~Eq.~\eqref{umut}] and that the critical gap is
asymptotically proportional to $\gamma$ [cf.~Eq.~\eqref{deltabeh}], we
may refine the definitions of $S_1$ and $S_2$, so that $S_1=w(t)
L/\gamma$ and $S_2 = \gamma t/L$, obtaining
\begin{eqnarray}
  \Upsilon = \frac{\gamma^2t_s}{L^2}, \qquad  \Omega = \frac{t}{\sqrt{t_s}}.
\label{scalvars2}
\end{eqnarray}

Assuming that the initial value $w_i$ remains fixed in the large-$L$
and large-$t_s$ limit, keeping $\Upsilon$ and $\Omega$ fixed, the
out-of-equilibrium KZ FSS of the fermionic correlations is
given by~\cite{RV-21}
\begin{equation}
  G_\#(t,x,t_s,w_i,\gamma,L) \approx L^{-2y_c}
  \widehat{\cal G}_\#(X,\Upsilon,\Omega), \label{KZscaG}
\end{equation}
where $y_c=1/2$ and $X=x/L$. Thus, the scaling behavior turns out to
be independent of the initial value $w_i$. The scaling functions $
\widehat{\cal G}_\#$ are expected to be universal and, in particular,
independent of $\gamma$, apart from a possible multiplicative factor.

%%%%%%%%%%%%%%%%%%%%%%%%%%%%%%%%%%%%%%%%%%%%%%%%%%%%%%%%%%%%%%%%%%%%%%%%
\begin{figure}[!t]
  \includegraphics*[scale=\graphicscale]{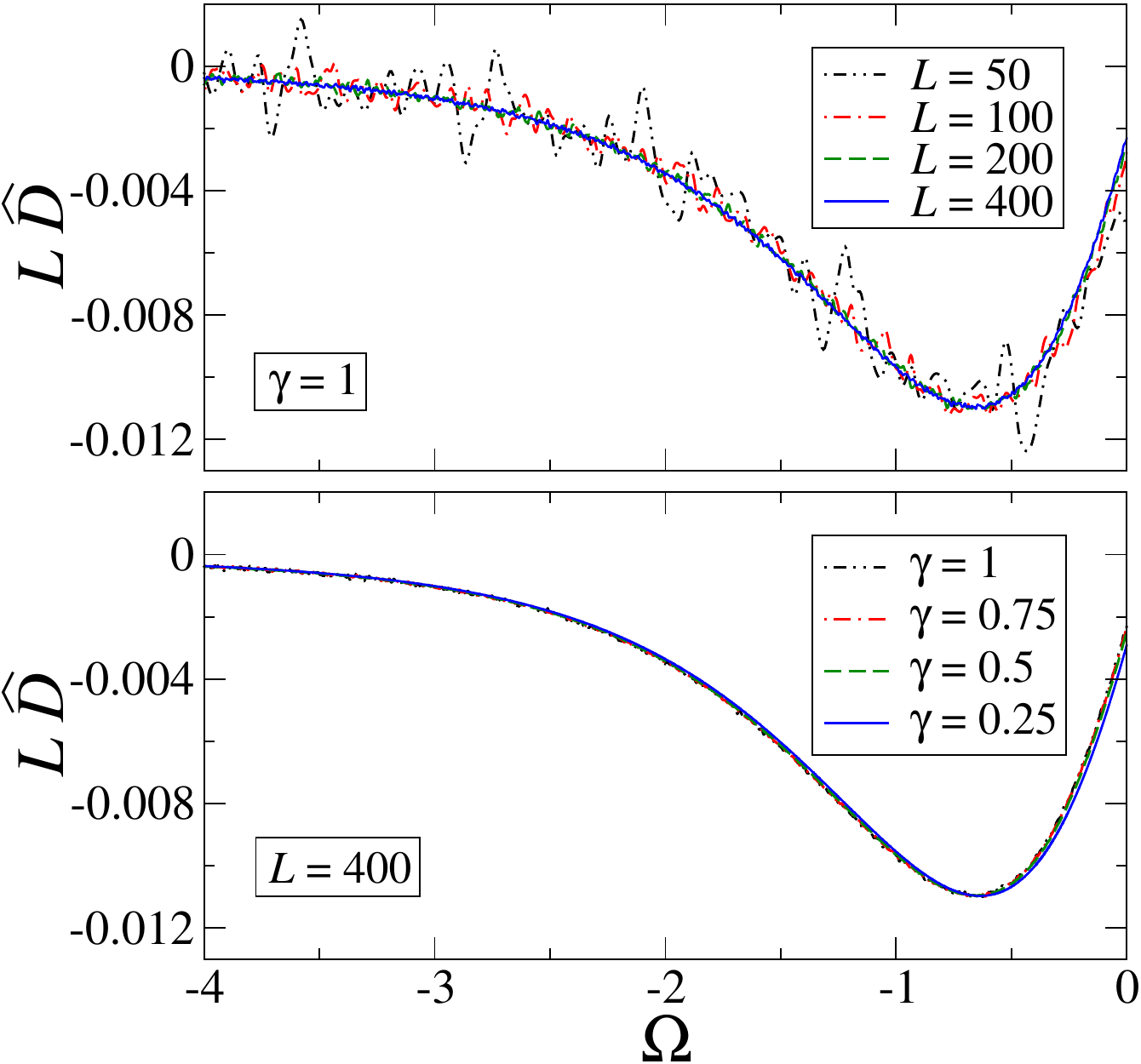}
  \caption{Deviation of the particle density from its equilibrium
    value $\widehat D$, along a KZ protocol from $w_i=0.5$ to
    $w_f=w_c=0$.  Data are shown against the rescaled time $\Omega$,
    in a window close to the ending point $\Omega=0$.  We fix
    $\Upsilon=0.1$ (analogous results are obtained for other values of
    $\Upsilon$). Upper panel: curves are for different values of $L$,
    while $\gamma = 1$ is kept fixed. They appear to approach an
    asymptotic curve, supporting Eq.~\eqref{KZscaD}.  Lower panel:
    curves are for different values of $\gamma$ while $L = 400$ is
    kept fixed, which are hardly distinguishable. Their agreement
    provides a strong evidence of universality of the scaling function
    $\widehat{\cal D}(\Upsilon,\Omega)$.}
  \label{KZbeh}
\end{figure}
%%%%%%%%%%%%%%%%%%%%%%%%%%%%%%%%%%%%%%%%%%%%%%%%%%%%%%%%%%%%%%%%%%%%%%%%

The out-of-equilibrium FSS of particle density following a KZ protocol
requires a more careful analysis, although its equilibrium behavior at
the transition is understood.  We may again consider the working
hypothesis that short-range fluctuations, responsible for the leading
equilibrium contributions to the particle density, get equilibrated
faster than the critical modes, so that the slow dynamics at the KZ
protocol, in the large $t_s$ limit keeping $\Upsilon$ and $\Theta$
fixed, can be considered as effectively adiabatic for them.
Therefore, the emerging out-of-equilibrium behavior in the KZ FSS
limit should be only related to the critical modes.  We again focus on
the deviation of the particle density from its equilibrium value
[cf.~Eq.~\eqref{diffet}],
\begin{equation}
  \widehat{D}(t,t_s,w_i,\gamma,L) =
          {1\over L} {\rm Tr}\,[\rho_\Psi(t) \,\hat{N}]
          - \varrho_e[w(t),\gamma,L],
          \label{kzddef}
\end{equation}
where $\varrho_e = \langle \Psi_{\rm GS}[w(t)] | \hat{n}_x | \Psi_{\rm
  GS}[w(t)]\rangle$ is the ground-state particle density at the
instantaneous value $w(t)$ and size $L$ of the system, whose
size-dependence at the transition was reported in Eq.~\eqref{fssrho}.
Then, analogously to Eq.~\eqref{KZscaG}, the above scenario naturally leads
to the conjecture 
\begin{equation}
  \widehat{D}(t,t_s,w_i,\gamma,L) \approx L^{-y_n} \widehat{\cal
    D}(\Upsilon,\Omega).  \label{KZscaD}
\end{equation}

As shown in Fig.~\ref{KZbeh}, this conjecture is fully supported by
our numerical results for different sizes $L$ (upper panel).  Besides
that, they also show that the asymptotic out-of-equilibrium scaling
behavior is universal, i.e., independent of the value of $\gamma$
(bottom panel).  In this respect we do not even apparently need a
multiplicative factor, while the $\gamma$ dependence in the
definitions of the scaling variables $\Upsilon$ and $\Omega$
[cf.~Eq.~\eqref{scalvars2}] already takes correctly into account their
nonuniversal normalizations.

\section{Conclusions}
\label{conclu}

We have addressed the equilibrium (ground-state) and
out-of-equilibrium behaviors of the particle density in many-body
systems undergoing quantum transitions driven by the chemical
potential, which arise from a nontrivial interplay between noncritical
short-range and critical long-range quantum fluctuations.  To study
this issue, we consider the fermionic Kitaev chain~\cite{Kitaev-01},
as a paradigmatic model for quantum transitions driven by the chemical
potential, for which very accurate numerical calculations, and thus
checks of the scaling ansatz, can be performed up to $O(10^4)$ sites.

The equilibrium behavior of the particle density (or equivalently the
transverse magnetization in the quantum $XY$ chain) is related to the
derivative of the free-energy density with respect to the
chemical-potential parameter $\mu$.  Its behavior at continuous
quantum transitions driven by $\mu$ is generally dominated by
contributions arising from short-range fluctuations, which may be
interpreted as mixings with the identity operators within CFT
frameworks~\cite{CHPV-02}. In particular, within the paradigmatic
fermionic Kitaev wires, they appear as a regular background term in
the free-energy density (actually this generally occurs at continuous
quantum transitions~\cite{CPV-14, RV-21}), and also logarithmic terms
arising from peculiar resonances~\cite{Wegner-76} between operators of
the {\em energy} and {\em identity} CFT families~\cite{CPV-14,
  CHPV-02}.  These contributions hide the genuine scaling behavior
arising from the long-range critical modes, making the particle
density nonoptimal to study the universal critical properties at
continuous quantum transitions.  The absence of an asymptotic
universal equilibrium scaling makes the out-of-equilibrium behavior
unclear. To study this issue, we focus on two simple dynamic
protocols: (i) instantaneous soft quenches, where the
chemical-potential is instantaneously changed from $\mu\neq\mu_c$
within the critical regime to its critical value $\mu_c$, and (ii)
quasi-adiabatic protocols, where $\mu$ gets slowly and linearly
changed from $\mu$ to $\mu_c$.

After subtracting its infinite-volume value at the critical point, the
subtracted particle density $D$ [cf.~Eq.~\eqref{diffet}] shows an
out-of-equilibrium FSS along the quantum evolution after a soft quench
of $\mu$.  The resulting asymptotic out-of-equilibrium FSS,
$D(t,w,\gamma,L) \approx L^{-y_n} {\cal D}(\Phi,\Theta)$, is
controlled by the RG dimension $y_n$ of the particle-density operator
$\hat{n}_x$ and is defined in the large-$L$ limit keeping the scaling
variables $\Phi\sim (\mu-\mu_c) L^{1/\nu}$ (associated with the
chemical-potential parameter) and $\Theta \sim t/L^z$ (associated with
the time interval from the quench) fixed.  This dynamic FSS appears
analogous to that of other observables possessing an asymptotic
equilibrium FSS~\cite{PRV-18, RV-21}, such as that of the fermionic
correlations [cf.~Eq.~\eqref{gxscaout}].  However, unlike them, the
scaling functions are now characterized by a logarithmic divergence in
the $\Theta\to 0$ limit, which is somehow related to the logarithmic
singularity at equilibrium [cf.~Eq.~\eqref{asyde}], which must be
somehow reconstructed in the $\Theta\to 0$ limit.  Analogous results
are obtained in the thermodynamic limit, in terms of the time scaling
variable $\theta \sim \Phi \Theta \sim (\mu-\mu_c) t$ remaining after
the thermodynamic infinite-size limit.

Within the out-of-equilibrium FSS framework, we also spotlight
logarithmic divergences at finite values of $\Theta$ (both for the
subtracted particle density and for the fermionic correlations), which
can be related to revival phenomena in finite-size systems, already
observed in various contexts (see, e.g., Refs.~\onlinecite{IR-11,
  RI-11, BRI-12, HHH-12, KLM-14, Cardy-14, JH-17, MAC-20, RV-20}).

Then we consider quasi-adiabatic KZ-like protocols, where the chemical
potential is slowly changed at the quantum transition, with a large
time scale (see Sec.~\ref{KZprot}).  Along these protocols, the
critical large-scale modes are generally unable to equilibrate the
long-distance critical correlations emerging at the transition point,
giving rise to peculiar out-of-equilibrium KZ FSS
behaviors~\cite{RV-21}.  However, the case of the particle density is
again particular, because its equilibrium behavior is dominated by the
short-range contributions. Under the assumption that the time scale of
the changes of the short-range modes, and therefore of the changes of
their contributions, is much smaller than that driving the critical
modes, the slow dynamics is expected to disentangle the effects of the
out-of-equilibrium critical modes from those associated with the
short-range modes.  This scenario leads us to conjecture that the
quasi-adiabatic KZ dynamics in the out-of-equilibrium FSS limit is
effectively adiabatic with respect to the short-range modes, and thus
the out-of-equilibrium behavior is only associated with the critical
modes that give rise to a well defined out-of-equilibrium FSS. This is
indeed observed when looking at the difference between the
out-of-equilibrium particle density and its equilibrium value at the
instantaneous value of the chemical-potential parameter
[cf.~Eq.~\eqref{kzddef}], which shows an out-of-equilibrium KZ FSS
analogous to the observables defined from the fermionic correlations,
whose equilibrium FSS is not affected by short-range contributions.

We remark that analogous out-of-equilibrium scaling behaviors, at both
instantaneous quenches and quasi-adiabatic protocols, are expected at any
quantum transition, when considering the behavior of observables
related to the derivative of the free-energy density with respect to
the Hamiltonian parameter that drives the quantum transition
preserving the symmetry (such as the transverse magnetization at the
quantum transitions of $d$-dimensional quantum Ising systems, or the
square angular momentum at the quantum transitions of $d$-dimensional
quantum rotor models~\cite{Sachdev-book,RV-21}).

We finally mention that recent theoretical proposals~\cite{SD-12,
  PD-23}, as well as experimental attempts to realize fermionic Kitaev
wires, by means of quantum dots~\cite{Dvir-etal-23}, integrated
circuits~\cite{IYMHYE-23}, or even quantum
computers~\cite{Huang-etal-21, SBEP-21, Rancic-22, Mi_etal-22}, have
been put forward with the purpose to manipulate Majorana zero modes.
Since, in this context, particle density measurements should be quite
accessible, we believe there could be the possibility to study its
out-of-equilibrium behavior, thus providing a further effective probe
of the universal features at quantum transitions (unlike its
equilibrium behavior, that is essentially dominated by the
nonuniversal short-range fluctuations).  More in general, the main
features of the dynamic scaling behaviors of the particle density for
the fermionic Kitaev wire are expected to extend to generic quantum
transitions driven the chemical potential, when the particle-number
operator in not conserved.  Further theoretical as well as
experimental studies could help in validating this scenario.

\acknowledgments

We thank Claudio Bonati for interesting and useful discussions.

\end{document}